\documentclass[english,aps,manuscript]{revtex4}
\usepackage[T1]{fontenc}
\usepackage[latin9]{inputenc}
\usepackage{amstext}
\usepackage{graphicx}
\usepackage{esint}

\makeatletter
\@ifundefined{textcolor}{}
{%
 \definecolor{BLACK}{gray}{0}
 \definecolor{WHITE}{gray}{1}
 \definecolor{RED}{rgb}{1,0,0}
 \definecolor{GREEN}{rgb}{0,1,0}
 \definecolor{BLUE}{rgb}{0,0,1}
 \definecolor{CYAN}{cmyk}{1,0,0,0}
 \definecolor{MAGENTA}{cmyk}{0,1,0,0}
 \definecolor{YELLOW}{cmyk}{0,0,1,0}
 }

\makeatother

\usepackage{babel}
\begin{document}

\title{Complex saddle points in QCD at finite temperature and density}

\author{Hiromichi Nishimura}

\address{Faculty of Physics, University of Bielefeld, D-33615 Bielefeld, Germany}

\author{Michael C. Ogilvie and Kamal Pangeni}

\address{Washington University, St. Louis, MO 63130 USA}

\date{01/30/14}
\begin{abstract}
The sign problem in QCD at finite temperature and density leads naturally
to the consideration of complex saddle points of the action or effective
action. The global symmetry $\mathcal{CK}$ of the finite-density
action, where $\mathcal{C}$ is charge conjugation and $\mathcal{K}$
is complex conjugation, constrains the eigenvalues of the Polyakov
loop operator $P$ at a saddle point in such a way that the action
is real at a saddle point, and net color charge is zero. The values
of $\mathrm{Tr}_{F}P$ and $\mathrm{Tr}_{F}P^{\dagger}$ at the saddle
point are real but not identical, indicating the different free energy
cost associated with inserting a heavy quark versus an antiquark into
the system. At such complex saddle points, the mass matrix associated
with Polyakov loops may have complex eigenvalues, reflecting oscillatory
behavior in color-charge densities. We illustrate these properties
with a simple model which includes the one-loop contribution of gluons
and two flavors of massless quarks moving in a constant Polyakov loop
background. Confinement-deconfinement effects are modeled phenomenologically
via an added potential term depending on the Polyakov loop eigenvalues.
For sufficiently large temperature $T$ and quark chemical potential
$\mu$, the results obtained reduce to those of perturbation theory
at the complex saddle point. These results may be experimentally relevant
for the Compressed Baryonic Matter (CBM) experiment at FAIR.
\end{abstract}


\maketitle
The sign problem is a fundamental issue in the study of QCD at finite
density, manifesting as complex weights in the path integral that
make lattice simulations extremely problematic \cite{deForcrand:2010ys,Gupta:2011ma,Aarts:2013bla}.
However, the sign problem also appears in analytical calculations
of mean-field type. Here we show that the sign problem can be solved
in such calculations provided a fundamental symmetry of finite-density
models, $\mathcal{CK}$ symmetry, is respected. This leads to the
analytic continuation of Polyakov loop eigenvalues into the complex
plane from the unit circle. Our results are complementary to recent
work on simulations of lattice field theories at finite densities
using the theory of Lefschetz thimbles \cite{Cristoforetti:2012uv,Fujii:2013sra,Cristoforetti:2013qaa,Cristoforetti:2013wha,Cristoforetti:2014gsa,Cristoforetti:2014waa},
and give an indication of how analytical and simulation results might
be combined to give a comprehensive picture of gauge theories at finite
density.

Let us consider an $SU(N)$ gauge theory coupled to fermions in the
fundamental representation. It is well known that the log of the fermion
determinant, $\log\det\left(\mu,A\right)$, which is a function of
the quark chemical potential $\mu$ and the gauge field $A$, can
be formally expanded as a sum over Wilson loops with real coefficients.
For a gauge theory at finite temperature, the sum includes Wilson
loops that wind nontrivially around the Euclidean timelike direction;
Polyakov loops, also known as Wilson lines, are examples of such loops.
At $\mu=0$, every Wilson loop $\mathrm{Tr}_{F}W$ appearing in the
expression for the fermion determinant is combined with its conjugate
$\mathrm{Tr}_{F}W^{\dagger}$ to give a real contribution to path
integral weighting. This can be understood as a consequence of charge
conjugation invariance $\mathcal{C},$ which acts on the gauge field
as

\begin{equation}
\mathcal{C}:\, A_{\mu}\rightarrow-A_{\mu}^{t}
\end{equation}
and thus exchanges $\mathrm{Tr}_{F}W$ and $\mathrm{Tr}_{F}W^{\dagger}.$
When $\mu\ne0$, Wilson loops with nontrivial winding number $n$
in the $x_{4}$ direction receive a weight $e^{n\beta\mu}$ while
the conjugate loop is weighted by $e^{-n\beta\mu}$. Thus it is seen
that these loops break charge conjugation invariance when $\mu\ne0$.
However, $\mathrm{Tr}_{F}W$ transforms into itself under the combined
action of $\mathcal{CK}$, where $\mathcal{K}$ is the fundamental
antilinear operation of complex conjugation. Thus the theory is invariant
under $\mathcal{CK}$ even in the case $\mu\ne0$. For fermions, $\mathcal{CK}$
symmetry implies the well-known relation $\det\left(-\mu,A_{\mu}\right)=\det\left(\mu,A_{\mu}\right)^{*}$
for Hermitian $A_{\mu}$, but can be used with bosons as well as fermions.

Given the existence of the symmetry $\mathcal{CK}$ at finite density,
we wish to ensure that, in the absence of spontaneous symmetry breaking,
calculational methods of all types respect the symmetry. For perturbative
or mean-field type calculations, this leads naturally to the consideration
of complex but $\mathcal{CK}$-symmetric saddle points for some effective
potential at finite temperature and density $V_{eff}$ , such that
the free energy density is given by the value of $V_{eff}$ at the
dominant saddle point. Such saddle points have been seen before in
finite-density calculations \cite{Hands:2010vw,Hands:2010zp,Hollowood:2011ep,Hollowood:2012nr}.
A field configuration is $\mathcal{CK}$ symmetric if $-A_{\mu}^{\dagger}$
is equivalent to $A_{\mu}$ under a gauge transformation. For such
a field configuration, it is easy to see that every Wilson loop is
real and thus $\det\left(\mu,A_{\mu}\right)$ is real and positive.

Let us consider the Polyakov loop $P$ associated with some particular
field configuration that is $\mathcal{CK}$ symmetric. We can transform
to Polyakov gauge where $A_{4}$ is diagonal and time independent,
and work with the eigenvalues $\theta_{j}$ defined by
\begin{equation}
P\left(\vec{x}\right)=\mbox{diag}\left[e^{i\theta_{1}\left(\vec{x}\right)},\cdots,\, e^{i\theta_{N}\left(\vec{x}\right)}\right]
\end{equation}
where the $\theta_{j}$'s are complex but satisfy $\sum_{j}\theta_{j}=0$.
Because we are primarily interested in constant saddle points, we
suppress the spatial dependence hereafter. Invariance under $\mathcal{CK}$
means that the set $\left\{ -\theta_{j}^{*}\right\} $ is equivalent
to the $\left\{ \theta_{j}\right\} $ although the eigenvalues themselves
may permute. In the case of $SU(3)$, we may write this set uniquely
as
\begin{equation}
\left\{ \theta-i\psi,-\theta-i\psi,2i\psi\right\} .
\end{equation}
This parametrizes the set of $\mathcal{CK}$-symmetric $SU(3)$ Polyakov
loops. Notice that both
\begin{equation}
\mathrm{Tr_{\mathnormal{F}}}P=e^{\psi}2\cos\theta+e^{-2\psi}
\end{equation}
 and
\begin{equation}
\mathrm{Tr_{\mathnormal{F}}}P^{\dagger}=e^{-\psi}2\cos\theta+e^{2\psi}
\end{equation}
are real, but they are equal only if $\psi=0$. In the usual interpretation
of the Polyakov loop expectation value, this implies that the free
energy change associated with the insertion of a fermion is different
from the free energy change associated with its antiparticle. It is
easy to check that the trace of all powers of $P$ or $P^{\dagger}$
are all real, and thus all group characters are real as well. This
parametrization represents a generalization of the Polyakov loop parametrization
used in the application of mean-field methods to confinement, $e.g.$,
in PNJL models \cite{Fukushima:2003fw} or in gauge theories with
double-trace deformations \cite{Myers:2007vc,Unsal:2007vu}. This
parametrization can be generalized to include finite-density models
for arbitrary $N$.

The existence of complex $\mathcal{CK}$-symmetric saddle points provides
a fundamental approach to non-Abelian gauge theories that is similar
to the heuristic introduction of color chemical potentials, and naturally
ensures the system has zero color charge, \emph{i.e.}, all three charges
contribute equally \cite{Buballa:2005bv}. In the case of $SU(3)$,
extremization of the thermodynamic potential with respect to $\theta$
leads to the requirement $\left\langle n_{r}\right\rangle -\left\langle n_{g}\right\rangle =0$
where $\left\langle n_{r}\right\rangle $ is red color density, including
the contribution of gluons. Similarly, extremization of the thermodynamic
potential with respect to $\psi$ leads to $\left\langle n_{r}\right\rangle +\left\langle n_{g}\right\rangle -2\left\langle n_{b}\right\rangle =0$.
Taken together, these two relations imply that $\left\langle n_{r}\right\rangle =\left\langle n_{g}\right\rangle =\left\langle n_{b}\right\rangle $.

We demand that any saddle point solution be stable to constant, real
changes in the Polyakov loop eigenvalues, corresponding for $SU(3)$
to constant real changes in $A_{4}^{3}$ and $A_{4}^{8}$. Consider
the $\left(N-1\right)\times\left(N-1\right)$ matrix $M_{ab}$, defined
in Polyakov gauge as 
\begin{equation}
M_{ab}\equiv g^{2}\frac{\partial^{2}V_{eff}}{\partial A_{4}^{a}\partial A_{4}^{b}}.
\end{equation}
The stability criterion is then that the eigenvalues of $M$ must
have positive real parts. At $\mathcal{CK}$-symmetric saddle points,
the eigenvalues will be either real or part of a complex conjugate
pair. In the case of $SU(3),$ the matrix $M$ may also be written
in terms of derivatives with respect to $\theta$ and $\psi$ as
\begin{equation}
M=\frac{g^{2}}{T^{2}}\left(\begin{array}{cc}
\frac{1}{4}\frac{\partial^{2}V_{eff}}{\partial\theta^{2}} & \frac{i}{4\sqrt{3}}\frac{\partial^{2}V_{eff}}{\partial\theta\partial\psi}\\
\frac{i}{4\sqrt{3}}\frac{\partial^{2}V_{eff}}{\partial\theta\partial\psi} & \frac{-1}{12}\frac{\partial^{2}V_{eff}}{\partial\psi^{2}}
\end{array}\right).
\end{equation}
This stability criterion generalizes the stability criterion used
previously for color chemical potentials, which was $\partial^{2}V_{eff}/\partial\psi^{2}<0$.
Crucially, the mass matrix $M$ is invariant under $M^{*}=\sigma_{3}M\sigma_{3}$,
which is a generalized parity-time $\left(\mathcal{PT}\right)$ symmetry
transformation \cite{Bender:1998ke,Meisinger:2012va}. It is easy
to see that this relation implies that $M$ has either two real eigenvalues
or a complex eigenvalue pair.

We first illustrate the working of $\mathcal{CK}$ symmetry using
the well-known one-loop expressions for the effective potential of
particles moving in a constant background Polyakov loop. The one-loop
contribution to the effective potential of $N_{f}$ flavors of fundamental
fermions moving in a background gauge field $A$ is given by
\begin{equation}
\beta\mathcal{V}V_{eff}^{f}=-N_{f}\log\left[\det\left(\mu,A\right)\right]
\end{equation}
where $\det$ again represents the functional determinant of the Dirac
operator and $\beta\mathcal{V}$ is the volume of spacetime. A compact
expression for the effective potential of massless fermions when the
eigenvalues of $P$ are complex was derived using zeta function methods
in \cite{KorthalsAltes:1999cp}. It will be useful in what follows
to repeat their derivation and check the reality of $V_{eff}^{f}$
for $\mathcal{CK}$-symmetric backgrounds. Our starting point is the
finite-temperature contribution to the effective potential of a single
Dirac fermion in a $U(1)$ background Polyakov loop characterized
by an angle $\theta$:

\begin{equation}
v_{f}(\theta)=-2T\int\frac{d^{3}k}{\left(2\pi\right)^{3}}\left\{ \log\left[1+e^{-\beta\omega_{k}+i\theta}\right]+\log\left[1+e^{-\beta\omega_{k}-i\theta}\right]\right\} .
\end{equation}
Setting the fermion mass to zero, and expanding the logarithms, we
obtain
\begin{equation}
v_{f}(\theta)=-\frac{4T^{4}}{\pi^{2}}\sum_{n=1}^{\infty}\frac{\left(-1\right)^{n+1}}{n^{4}}\cos\left(n\theta\right).
\end{equation}
After expanding the cosine and interchanging the order of summation,
we get

\begin{equation}
v_{f}(\theta)=-\frac{4T^{4}}{\pi^{2}}\sum_{m=0}^{\infty}\frac{(-1)^{m}\theta^{2m}\eta(4-2m)}{(2m)!}
\end{equation}
where $\eta$ is the Dirichlet eta function. Only the first three
terms of the expansion are nonzero, and we arrive at 
\begin{equation}
v_{f}(\theta)=-\frac{4T^{4}}{\pi^{2}}\left(\frac{\theta^{4}}{48}-\frac{\pi^{2}\theta^{2}}{24}+\frac{7\pi^{4}}{720}\right).
\end{equation}
This expression is valid provided $Re\left[\theta\right]\in\left(-\pi,\pi\right)$.
In general, if $\theta$ is complex, so is $v_{f}$. However, the
free energy of quarks in a $\mathcal{CK}$-symmetric background Polyakov
loop is always real. For $SU(3)$, we have
\begin{equation}
V_{f}(\theta,\psi,\text{T},\mu)=N_{f}\left(v_{f}\left(\theta-i\psi-\frac{i\mu}{T}\right)+v_{f}\left(-\theta-i\psi-\frac{i\mu}{T}\right)+v_{f}\left(2i\psi-\frac{i\mu}{T}\right)\right)
\end{equation}
 which is explicitly real. For two massless flavors, the result is
\begin{eqnarray}
V_{f}(\theta,\psi,\text{T},\mu) & = & -\frac{\mu^{4}}{2\pi^{2}}+T^{2}\left(-\mu^{2}+\frac{2\theta^{2}\mu^{2}}{\pi^{2}}-\frac{6\mu^{2}\psi^{2}}{\pi^{2}}\right)+\frac{4T^{3}\left(\theta^{2}\mu\psi+\mu\psi^{3}\right)}{\pi^{2}}\nonumber \\
 &  & +\frac{T^{4}\left(-7\pi^{4}+20\pi^{2}\theta^{2}-10\theta^{4}-60\pi^{2}\psi^{2}+60\theta^{2}\psi^{2}-90\psi^{4}\right)}{30\pi^{2}}.
\end{eqnarray}

Because we are interested in the analytic continuation of Polyakov
loop eigenvalues into the complex plane, we need expressions for the
gauge bosons as well as for fermions. Our starting point in this case
is
\begin{equation}
v_{b}(\theta)=2T\int\frac{d^{3}k}{\left(2\pi\right)^{3}}\left\{ \log\left[1-e^{-\beta\omega_{k}+i\theta}\right]+\log\left[1-e^{-\beta\omega_{k}-i\theta}\right]\right\} 
\end{equation}
which represents the one-loop of two gauge bosons of opposite $U(1)$
charge. A naive repetition of the zeta-function argument that was
successful for fermions fails for bosons even in the case where $\theta$
is real. The expansion of $v_{b}\left(\theta\right)$ around $\theta=0$
is invalid, because the final result is only valid for $0\le Re\left[\theta\right]\le2\pi$.
In contrast, the fermionic expression is valid for $-\pi\le Re\left[\theta\right]\le\pi$.
However, it is possible to define the bosonic analog of $v_{f}\left(\theta\right)$
by \cite{Meisinger:2001fi}
\begin{equation}
v_{b}\left(\theta\right)=-v_{f}\left(\theta-\pi\right)
\end{equation}
 and this leads to the correct expression when $Re\left[\theta\right]\in\left(0,2\pi\right)$:
\begin{equation}
v_{b}\left(\theta\right)=-\frac{\left(-15\theta^{4}+60\pi\theta^{3}-60\pi^{2}\theta^{2}+8\pi^{4}\right)T^{4}}{180\pi^{2}}.
\end{equation}
As in the fermionic case, $v_{b}\left(\theta\right)$ is generally
complex if $\theta$ is complex. However, the one-loop gluonic contribution
for a $\mathcal{CK}$-symmetric Polyakov loop background, given by
\begin{equation}
V_{g}(P)=v_{b}(0)+\text{\ensuremath{v_{b}}}(2\theta)+\text{\ensuremath{v_{b}}}(\theta+i3\psi)+\text{\ensuremath{v_{b}}}(\theta-i3\psi)
\end{equation}
 is real. Explicitly, we have for $SU(3)$
\begin{equation}
V_{g}(P)=\frac{T^{4}\left(135\left(\theta^{2}-3\psi^{2}\right)^{2}+180\pi^{2}\left(\theta^{2}-3\psi^{2}\right)+60\pi\theta\left(27\psi^{2}-5\theta^{2}\right)-16\pi^{4}\right)}{90\pi^{2}}
\end{equation}
which is real. Note that the valid range of $\theta$ is $\left(0,\pi\right)$
due to the appearance of $2\theta$ as an eigenvalue in the adjoint
representation. The one-loop effective potential is simply the sum
of $V_{g}(\theta)$ and $V_{f}(\theta)$. As is the case when $\mu=0$,
the dominant saddle point remains at $\theta=0$ when $\mu\ne0$.

We now consider a simple phenomenological model that combines the
one-loop result with the effects of confinement for the case of $SU(3)$
gauge bosons and two flavors of massless fermions at finite temperature
and density. The model is described by an effective potential which
is the sum of three terms: 
\begin{equation}
V_{eff}(P)=V_{g}(P)+V_{f}(P)+V_{d}(P)
\end{equation}
where $V_{g}(P)+V_{f}(P)$ is the one-loop effective potential for
gluons and quarks given above and $V_{d}(P)$ is an additional term
that favors the confined phase \cite{Meisinger:2001cq,Myers:2007vc,Unsal:2008ch,Ogilvie:2012is}.
There are two different points of view that can be taken on this model.
In one view, $V_{d}(P)$ represents a deformation of the original
model. In typical applications, the temperature $T$ is taken to be
large such that perturbation theory is reliable in the chromoelectric
sector because the running coupling $g^{2}\left(T\right)$ is small.
The deformation term is taken to respect center symmetry and is used
to move between the confined and deconfined phases in a controlled
way. The gauge contribution $V_{g}(P)$ favors the deconfined phase,
and in the pure gauge theory ($N_{f}=0$) the deconfinement transition
arises out of the competition between $V_{g}(P)$ and $V_{d}(P)$.
The confined phase arising in models of this type is known to be analytically
connected to the usual low-temperature confined phase of $SU(3)$
gauge theory \cite{Myers:2007vc}. This point of view emphasizes analytic
control at the price of deforming the original gauge theory by the
addition of $V_{d}(P)$. From the second point of view, the entire
model is phenomenological in nature, with the potential $V_{d}(P)$
models the unknown confining dynamics of the pure gauge theory. The
parameters of $V_{d}(P)$ are set to reproduce the deconfinement temperature
of the pure gauge theory, known from lattice simulations to occur
at $T_{d}\approx270\,\mbox{MeV}$. 

We will take the second point of view, using a simple expression for
$V_{d}(P)$ that reproduces much of the thermodynamic behavior seen
in lattice simulations of the pure gauge theory. The specific form
used is Model A of \cite{Meisinger:2001cq}, where the two terms $V_{g}(P)+V_{d}(P)$
can be written together as 
\begin{equation}
V_{A}\left(\theta\right)=-\sum_{j,k=1}^{N}\frac{1}{\pi^{2}}(1-\frac{1}{N}\delta_{jk})\left[-\frac{2\pi^{4}}{3\beta^{4}}B_{4}\left(\frac{\Delta\theta_{jk}}{2\pi}\right)-\frac{m^{2}\pi^{2}}{2\beta^{2}}B_{2}\left(\frac{\Delta\theta_{jk}}{2\pi}\right)\right]
\end{equation}
where $\Delta\theta_{jk}=\left|\theta_{j}-\theta_{k}\right|$ are
the adjoint Polyakov loop eigenvalues and $B_{j}$ is the $j$'th
Bernoulli polynomial. This expression gives a simple quartic polynomial
in the Polyakov loop eigenvalues and thus can be thought of as a form
of Landau-Ginsburg potential for the Polyakov loop eigenvalues. For
the $SU(3)$ parametrization used here, $V_{d}(P)$ takes the simple
form
\begin{equation}
V_{d}\left(P\right)=\frac{m^{2}T^{2}\left((2\pi-3\theta)^{2}-27\psi^{2}\right)}{6\pi^{2}}.
\end{equation}
The parameter $m$ controls the location of the deconfinement transition
in the pure gauge theory, and is set to $596\,\mbox{MeV}$. At low
temperatures, this term dominates the pure gauge theory effective
potential. The variable $\psi$ is zero, and $V_{d}\left(P\right)$
is minimized when $\theta=2\pi/3$. For this value of $\theta$, the
eigenvalues of $P$ are uniformly spaced around the unit circle, respecting
center symmetry, and $\mathrm{Tr}_{F}P=\mathrm{Tr}_{F}P^{\dagger}=0$.
As the temperature increases, $V_{g}\left(P\right)$ becomes relevant,
and gives rise to the deconfined phase where center symmetry is spontaneously
broken. The addition of light fundamental quarks via $V_{f}(P)$ explicitly
breaks center symmetry. For all nonzero temperatures, center symmetry
is broken and $\left\langle \mathrm{Tr}_{F}P\right\rangle \ne0$.
However, a remnant of the deconfinement transition remains in the
form of a rapid crossover from smaller value of $\mathrm{Tr}_{F}P$
to larger ones as $T$ and $\mu$ are varied. Note that this simple
model neglects both chiral symmetry breaking, relevant at low $T$
and low $\mu$ and the formation of a color superconductor, which
occurs at low $T$ and high $\mu$. Because the simplified model we
are using does not treat chiral symmetry breaking, it should not be
expected to reproduce exactly the features seen in lattice simulations.
Nevertheless, comparison with PNJL model results, \emph{e.g.}, \cite{Schaefer:2007pw},
show that the model is quantitatively similar to the behavior of models
with many more free parameters that include chiral symmetry effects.
In the model studied here, $\mathrm{Tr}_{F}P$ shows a slightly larger
initial rise in $\mathrm{Tr}_{F}P$ with temperature than does the
model studied in  Ref. \cite{Schaefer:2007pw}. This is consistent
with the role that chiral symmetry breaking plays in diminishing the
explicit breaking of $Z(3)$ symmetry by quarks. We plan to include
the effects of chiral symmetry breaking in a PNJL-type treatment in
a future work.

\begin{figure}
\includegraphics[width=5in]{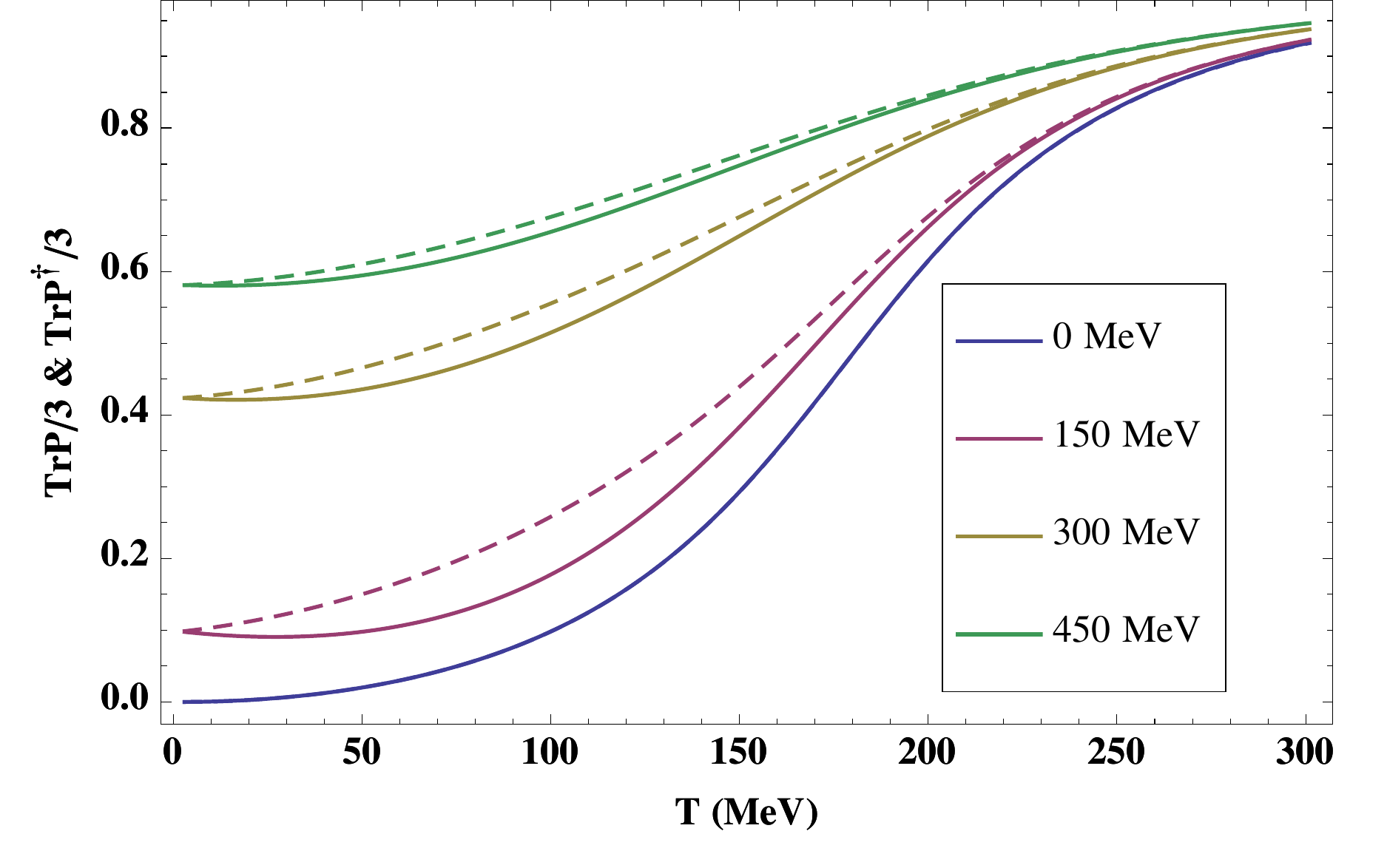}

\caption{\label{fig:PvsTconstMu}$\frac{1}{3}\mathrm{Tr}_{F}P$ (solid line)
and $\frac{1}{3}\mathrm{Tr}_{F}P^{\dagger}$ (dotted line) as functions
of $T$ at constant $\mu$ for values up to $\mu=450\ \mbox{MeV}$
with $N_{f}=2$.}
\end{figure}

\begin{figure}
\includegraphics[width=5in]{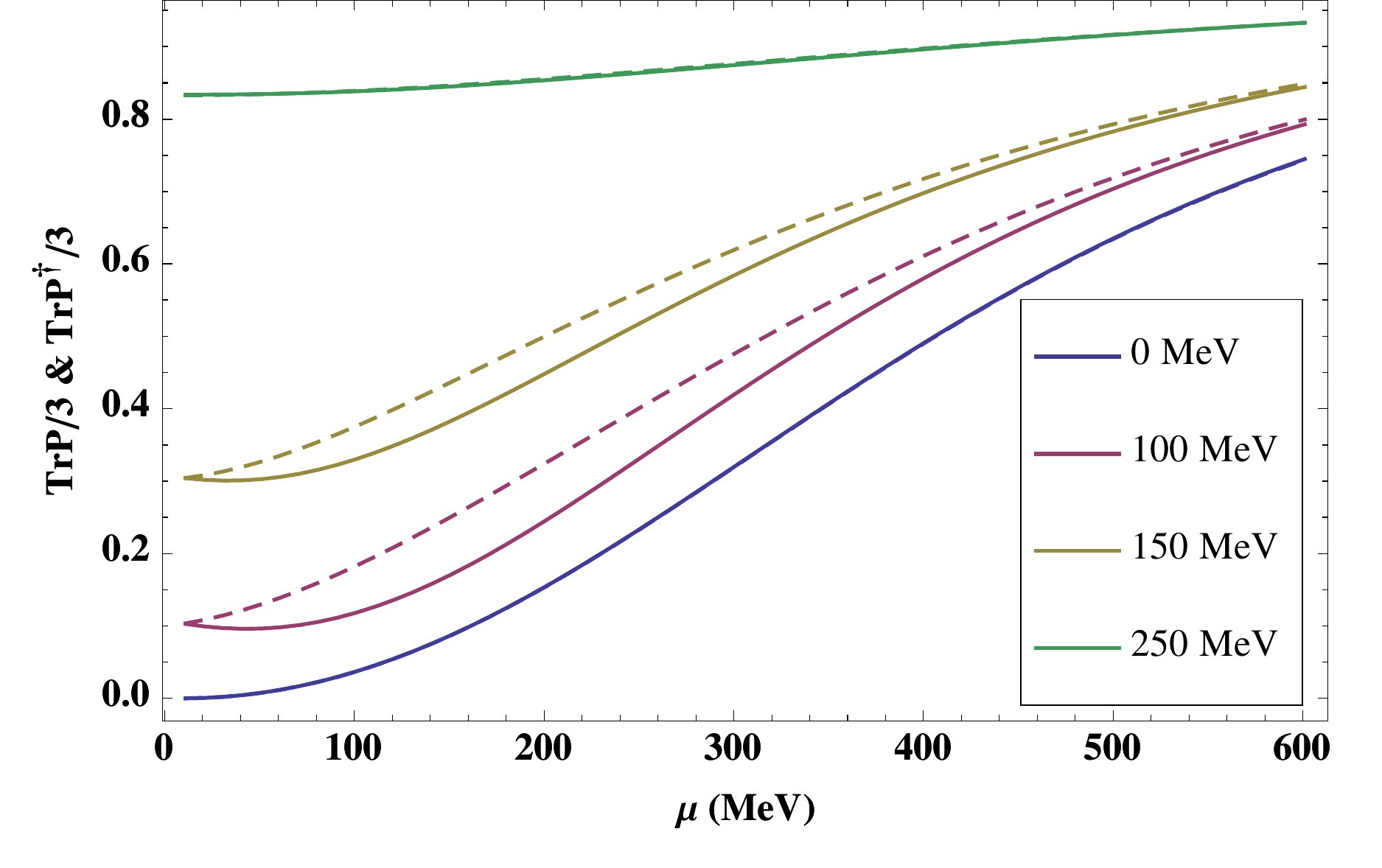}

\caption{\label{fig:PvsMuconstT}$\frac{1}{3}\mathrm{Tr}_{F}P$ (solid line)
and $\frac{1}{3}\mathrm{Tr}_{F}P^{\dagger}$ (dotted line) as a function
of $\mu$ at constant $T$ for values up to $T=250\,\mbox{MeV}$ with
$N_{f}=2$.}
\end{figure}

For a given $T$ and $\mu,$ the free energy and other thermodynamic
quantities are obtained from the saddle point of $V_{eff}(P)$. Figure
\ref{fig:PvsTconstMu} shows $\mathrm{Tr}_{F}P$ and $\mathrm{Tr}_{F}P^{\dagger}$
as functions of $T$ at constant $\mu$ for values up to $\mu=450\,\mbox{MeV}$
for two flavors of massless quarks. There is a rapid crossover in
$\mathrm{Tr}_{F}P$ and $\mathrm{Tr}_{F}P^{\dagger}$ for smaller
values of $\mu$ that becomes less dramatic as $\mu$ increases. There
is a small difference between $\mathrm{Tr}_{F}P$ and $\mathrm{Tr}_{F}P^{\dagger}$
in the crossover region when $\mu\ne0$, with $\mathrm{Tr}_{F}P^{\dagger}>\mathrm{Tr}_{F}P$,
indicating that it is easier to insert a heavy antiquark into the
system than a heavy quark. Figure \ref{fig:PvsMuconstT} shows $\mathrm{Tr}_{F}P$
and $\mathrm{Tr}_{F}P^{\dagger}$ as functions of $\mu$ at constant
$T$ for values up to $T=250\,\mbox{MeV}$. Similar behavior is obtained
as that shown in Fig. \ref{fig:PvsTconstMu}, but the crossover is
less abrupt and almost gone at $T=250\,\mbox{MeV}$. These results
are consistent with more complicated models that include the effects
of chiral symmetry and color superconductivity. 

\begin{figure}
\includegraphics[width=5in]{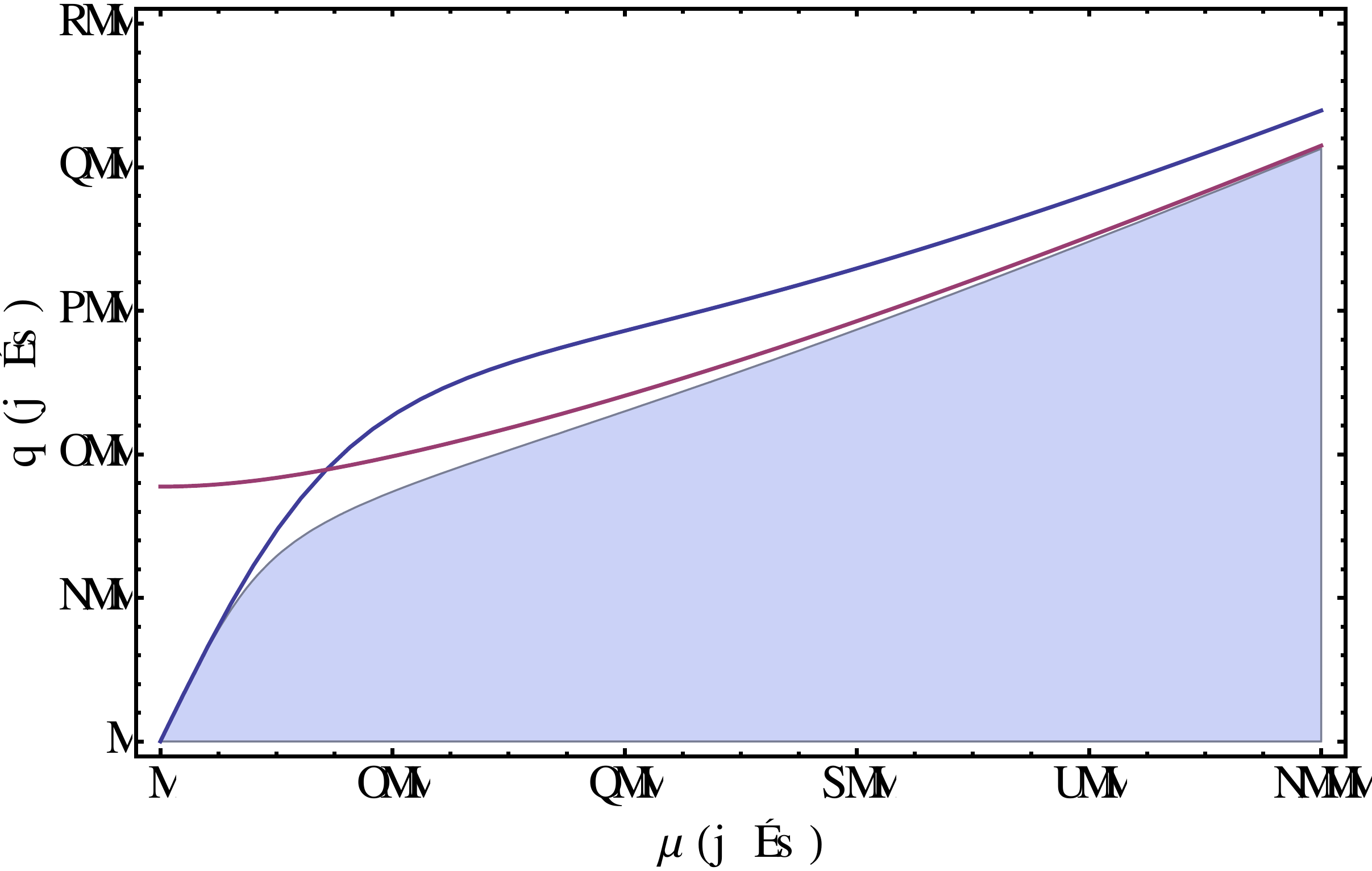}

\caption{\label{fig:Tmuline}The shaded region of the $\mu-T$ plane indicates
where the mass matrix is complex. High-$T$ and low-$T$ approximations
to the boundary are also shown.}
\end{figure}

As discussed above, the mass matrix associated with the fields $A_{4}^{3}$
and $A_{4}^{8}$ has the property that the eigenvalues are either
both real or form a complex conjugate pair. The most physically interesting
region for this model occurs when $T$ is larger than the $\mu=0$
crossover temperature. In this region, we can safely assume that chiral
symmetry is approximately restored, and the use of zero-mass quarks
is a reasonable approximation. Working in this region also excludes
color superconducting phases. Useful analytic results can be obtained
in the region $T,\mu\gg m$, where the saddle point is given approximately
by
\begin{eqnarray}
\theta & \approx & \frac{3m^{2}\pi}{8\pi^{2}T^{2}+6\mu^{2}}\\
\psi & \approx & \frac{9m^{4}\pi^{2}T\mu}{4\left(4\pi^{2}T^{2}+3\mu^{2}\right)^{3}}.
\end{eqnarray}
The corresponding mass matrix eigenvalues are given to order $m^{2}$
by
\begin{equation}
g^{2}\left(\frac{4T^{2}}{3}+\frac{\mu^{2}}{\pi^{2}}\right)+\frac{g^{2}m^{2}\left(9\mu^{2}-12\pi^{2}T^{2}\pm2\pi T\sqrt{9\pi^{2}T^{2}-12\mu^{2}}\right)}{4\pi^{2}\left(3\mu^{2}+4\pi^{2}T^{2}\right)},
\end{equation}
a formula that combines the one-loop formula for the screening masses
$g^{2}\left(4T^{2}/3+\mu^{2}/\pi^{2}\right)$ with corrections due
to the confining term $V_{d}$ in $V_{eff}$ that are proportional
to $m^{2}$. If $T>2\mu/\sqrt{3}\pi$, there are two real, nondegenerate
eigenvalues. However, if $T<2\mu/\sqrt{3}\pi$, the two eigenvalues
form a conjugate pair. The occurrence of such pairs is unusual in
a Euclidean field theory, and is associated with the sign problem
at finite density \cite{Meisinger:2012va}. In a $d$-dimensional
theory, a boson of mass $m_{B}$ contributes a term
\begin{equation}
\frac{1}{2}\int\frac{d^{d}k}{\left(2\pi\right)^{d}}\log\left[k^{2}+m_{B}^{2}\right]
\end{equation}
to the effective potential. A negative value for $m_{B}^{2}$ leads
to an imaginary part for the effective potential, indicating instability,
as would a complex value. However, in the case where there are complex
conjugate pairs of mass eigenvalues where $m_{B}^{2}=a\pm ib$, the
contribution of the two terms is now
\begin{equation}
\frac{1}{2}\int\frac{d^{d}k}{\left(2\pi\right)^{d}}\log\left[\left(k^{2}+a+ib\right)\left(k^{2}+a-ib\right)\right]=\frac{1}{2}\int\frac{d^{d}k}{\left(2\pi\right)^{d}}\log\left[\left(k^{2}+a\right)^{2}+b^{2}\right]
\end{equation}
which is always positive, and thus shows no instability. However,
correlation functions will in general exhibit modulated decay.

The occurrence of complex eigenvalues indicates periodic modulation
in the spatial decay of color-color density correlation functions
reminiscent of similar oscillations in density-density correlation
functions in liquids. Figure \ref{fig:Tmuline} shows the region where
the exact mass matrix is complex, along with high-$T$ and low-$T$
approximations to the boundary of the region. This result may depend
strongly on the choice of $V_{d}$, an issue we plan to address in
later work. Patel has suggested that a signal for such oscillatory
behavior might appear in baryon number correlators in heavy ion collisions
at RHIC and the LHC \cite{Patel:2011dp,Patel:2012vn}. Based on the
results reported here, it would be more natural to observe this behavior
in the CBM experiment at FAIR, but much work is needed to determine
a useful experimental signature.
\begin{acknowledgments}
HN was supported by the Sofja Kovalevskaja program of the Alexander
von Humboldt Foundation. MCO thanks the U.S. Department of Energy
for financial support, and the Department of Physics, University of
Washington for hospitality during the completion of this manuscript.
\end{acknowledgments}
\bibliographystyle{unsrt}
\bibliography{Complex_Saddles_Letter}

\end{document}